# Datacenter Energy Optimized Power Profiles

Sreedhar Narayanaswamy
Silicon Solutions Group
Nvidia
Santa Clara, USA
sreedharn@nvidia.com

Pratikkumar Patel
Software
Nvidia
Santa Clara, USA
pratikkumarp@nvidia.com

Ian Karlin
Accelerated Computing Product Group
Nvidia
Hillsboro, USA
ikarlin@nvidia.com

Apoorv Gupta
Software
Nvidia
Santa Clara, USA
apoorvg@nvidia.com

Sudhir Saripalli
Software
Nvidia
Santa Clara, USA
susaripalli@nvidia.com

Janey Guo
Silicon Solutions Group
Nvidia
Shanghai, China
janeyg@nvidia.com

## ABSTRACT

This paper presents datacenter power profiles, a new NVIDIA software feature released with Blackwell B200 GPUs, aimed at improving energy efficiency and/or performance. The initial feature provides coarse-grain user control for HPC and AI workloads leveraging hardware and software innovations for intelligent power management and domain knowledge of HPC and AI workloads. The resulting workload-aware optimization recipes maximize computational throughput while operating within strict facility power constraints. The phase-1 Blackwell implementation achieves up to 15% energy savings while maintaining performance levels above 97% for critical applications, enabling an overall throughput increase of up to 13% in a power-constrained facility.

## 1 Introduction

Energy efficiency is a cornerstone challenge for modern data centers supporting HPC and AI workloads. As computational demands increase exponentially while power infrastructure remains constrained, facility power has become a primary limit on computational throughput. For example, in 2022 Riken shut down parts of the Fugaku system due to rising power rates [8]. Therefore, strategies that maximize energy efficiency are essential not only for increasing a data center's total throughput but also for improving its sustainability and reducing operational costs without compromising performance.

While growing in importance, tuning applications for power and energy is becoming more complex. With each subsequent generation of GPUs, the number of settings that can be changed to improve energy efficiency grows. Leveraging these settings to tune applications for energy efficiency already requires expert-level understanding of workloads and GPU architecture with the challenge increasing with subsequent generations of hardware. While some work has aimed at standardizing interfaces [7] to these settings it does not make the tuning process easier.

Enabling non-experts to tune applications and achieve good energy efficiency with limited effort is an open challenge. Various tools, such as, EAR [3], Loadleveler [1] and GEOPM [5] take various approaches to this problem. However, each of these tools only adjust some of the available power settings and often do not incorporate any application specific knowledge into their tuning.

Our approach to this problem is a new out-of-box solution workload power profiles. Workload power profiles combine a high-level user interfaces, to enable ease of use by end users who are not experts in energy and power optimization, with a backend implementation that takes this input to adjust power settings for optimizing performance or energy efficiency of a job. Our initial implementation targets application classes based on hardware utilization characteristics. Application programmers, therefore, only need to know their application rather than hardware details to get started. Options for more fine-grain control based on other workload characteristics that can be found from profiling will be added over time.

Power profiles leverage application knowledge. For example, AI workloads extensively utilize 16-bit or lower precision tensor cores, while HPC applications primarily use FP32 or FP64 computations with minimal tensor core engagement. These computational differences result in applications operating at distinct points on the voltage-frequency curve, enabling optimization. In addition, we support hints to the application to enable further tuning based on computational intensity characteristics or I/O requirements, such as memory-bound vs. compute-bound and NVLINK heavy vs. NVLINK light.

This paper presents NVIDIA's Power profiles feature released with the Blackwell GPU [10], which creates workload-aware power profiles to maximize computational throughput within a fixed power envelope. This technology synthesizes hardware innovations, advanced software algorithms, and deep analysis of AI /HPC workload behavior. A high-level interface to these optimized profiles is delivered through the NVIDIA Mission Control framework for easy integration into data centers enabling significant energy savings with minimal user effort.

The innovations described in this paper are:

- A description of the software architecture of Workload Power Profiles and what it enables users and administrators to accomplish

- A simple extensible design enabling users access to power tuning features previously reserved for experts.

- Performance results on Blackwell B200 and Hopper H100 GPUs are presented showing up to a 15% improvement in application energy efficiency and a 13% improvement in data center throughput for HPC and AI applications.

- A roadmap of features that includes system level optimizations, and tools to ease energy and power optimization further.

In this paper we first describe the software architecture of Workload Power Profiles, their integration into NVIDIA mission control and the user interfaces and options available. We then present the performance of our tool and related work. Finally, we present conclusions and a roadmap of future directions.

## 2 Workload Power Profiles Implementation Work

This section describes how Workload Power Profiles are built and its four layers. The software stack is shown in Figure Fig. 2. The bottom layer is existing NVIDIA capabilities for controlling setting that impact energy efficiency. Layer 2 The Profile Abstraction Layer is where the tool is implemented. Layer 3 is where power profiles are exposed to users and administrators through lower-level interfaces. It also is where we implemented the capabilities to tunnel from the User Mode Driver (UMD) to the Kernel Mode Driver (KMD) through the NVIDIA Management Layer (NVML) interfaces. Layer 4 is a higher-level interface through NVIDIA mission control that simplifies the invoking and using power profiles.

Workload Power Profiles has two main modes Max-Q and Max-P. Borrowed from an aerospace term, Max-Q is a design approach for delivering energy efficiency into Nvidia product lines. It involves optimizing the hardware architecture, thermal and electrical components, firmware, drivers, control systems and more to enable maximum energy efficiency. Workload Power Profiles is one of the key initiatives in the larger Nvidia Datacenter Max-Q technology. Max-P is a second mode in the tool that aims to deliver maximum performance by optimizing the usage of energy and power under thermal designed power (TDP) constraints.

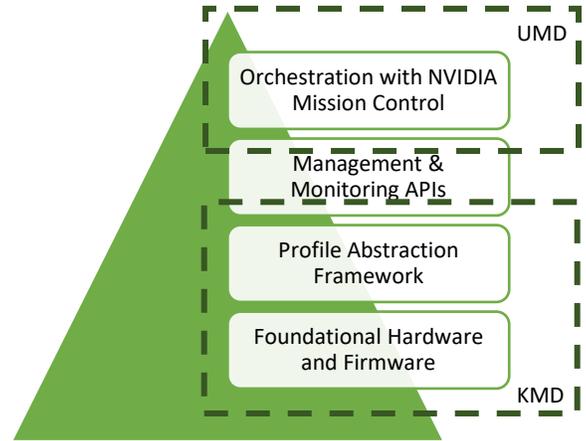

**Figure 1: The four layers of workload profiles software stack.**

**Layer 1: Foundational Hardware and Firmware.** This layer is all of the NVIDIA controls that enable users to impact power and energy usage. These include GPU and CPU clock controls, memory frequency settings (MCLK), total GPU power limits (TGP), dynamic voltage-frequency scaling (DVFS), and high-speed GPU interconnect (NVLink) power states. Most of these controls are exposed individually through tools like nvidia-smi, however the growing number of controls and the cross dependency between them requires expertise by end-users to co-optimize them for workloads. As a result, most users only adjust a handful of them.

Achieving optimal product energy efficiency relies on a multi-faceted approach, beginning with foundational hardware and architectural design. Key decisions made during this phase—from cache hierarchies and memory controllers to specific circuit choices—are critical. Furthermore, fine-tuning and calibrating settings across these hardware and architectural layers is essential for optimization.

This focus extends into the post-silicon stage, where often-overlooked manufacturing processes yield significant efficiency gains. Advanced techniques, such as strategic parts binning, dielet selection, and specific process choices, are crucial innovations that profoundly impact the final product's energy performance.

Sitting on top of the hardware at the lowest level of kernel model drivers (KMD) are our firmware algorithms and controls that enable key optimizations. Multiple control algorithms run to enforce thermal and power constraints in the chip and to comply with higher level power controls set by users and administrators including but are not limited to the Total GPU Power (TGP) Controller, Utilization Controller, DVFS with clock propagation, and VF Management. In addition, low level settings, such as, GPU interconnect NVLink power states, and Memory P-states are exposed.

This piece is often missed as the tight interaction between Hardware and the Firmware algorithms happens at layers below what users see. NVIDIA has spent a significant amount of effort in co-optimizing the interactions across multiple SW layers in the CUDA stack along with Hardware leading to the uplift in energy efficient gains compared to prior generations.

**Layer 2: The Profile Abstraction Layer.** This new layer, implemented within the NVIDIA KMD, introduces a critical abstraction. It shifts the paradigm from manipulating individual hardware knobs to using holistic power profiles. The abstraction layer uses firmware configuration tables where NVIDIA engineers define and tune specific combinations of the foundational controls from Layer 1. The result is a set of pre-configured, validated profiles (e.g., Max-Q for efficiency, Max-P for peak performance) optimized for targeted use cases like AI training or inference. We also define hardware feature level control knobs and embedded algorithms in this layer and some of the controls in this layer are unable to be exposed to users for security or intellectual property reasons and can only be accessed through the profiles tool.

The largest algorithmic innovation at this layer is the one that understands workload characteristics and enables specific control knobs that benefit energy efficiency for that class of workloads. It takes in user input of the job type, e.g. HPC, AI training, AI inference, properties of that job e.g. memory-bound, compute-bound, and user goals, e.g. Max-Q or Max-P. Then passes the optimal settings for this input to the performance mode infrastructure.

The performance mode infrastructure is a framework developed and designed to manage GPU configuration profiles. This infrastructure is composed of two primary blocks: performance modes and performance mode configurations. A performance mode is a high-level setting that maps to one or more specific performance mode configurations, each containing a defined value to be programmed for the device, such as a GPU. This modular design allows a team to create multiple performance modes that can share configurations.

The infrastructure supports the concept of coexisting performance modes, allowing multiple modes to be enabled simultaneously. To manage how these modes interact, the infrastructure incorporates an arbitration algorithm that utilizes priority and conflicting masks, enabling the definition of multiple modes that may or may not be able to coexist.

Arbitration is a key function of the infrastructure, using priority to resolve conflicts and overlaps. It is applied in two main scenarios. First, when two or more conflicting modes are engaged, the infrastructure selects the mode with the highest priority to be active. Second, when two non-conflicting modes both contain the same configuration knobs, the infrastructure chooses the knob value from the mode with the higher priority. Non-overlapping configurations from both active modes are merged.

This capability allows for flexible configurations, such as a user selecting a base mode like "Compute" and a modifier mode like "Max-P." The driver can then intelligently merge the configuration knobs from both modes to create a final, unified configuration set for the GPU. The configurations from the lower-priority, conflicting modes are then discarded and not applied. For example, if a "Compute" mode and a "Memory" mode are marked as conflicting, and both are enabled, the infrastructure will choose the one with the higher priority and ignore the configuration of the other. When this occurs, users are informed of the conflicts and made aware of which modes were used by the driver. In addition, users can query the tool to see the relative priority of all modes to understand the priority order of how conflicts are resolved.

**Layer 3: Management and Monitoring APIs.** The third layer exposes power profiles to users through extensions to NVIDIA's established data center toolset. Power profiles are exposed both through NVIDIA System Management Interface (nvidia-smi) and the Data Center GPU Manager (DCGM). These tools provide command line and application programmatic interfaces that allow users, scripts, and higher-level management software to query available profiles and apply them to specific GPUs. This capability is implemented using the Nvidia management library (NVML) APIs that enable these tools to tunnel from the user mode driver (UMD) to kernel mode driver (KMD) and access the capabilities and APIs implemented in layer 2.

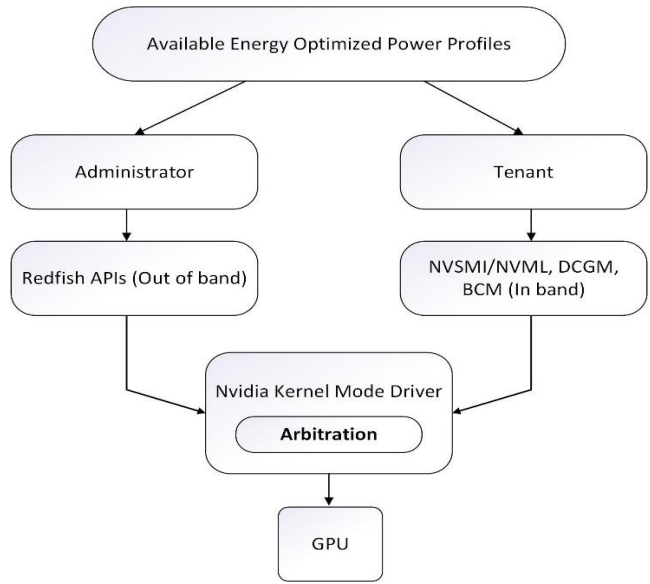

**Figure 2: Control Flow Diagram for the management and monitoring APIs.**

As depicted in Fig. 2, there are two primary user types: Administrators e.g. HPC system admins, and Cloud Service Providers (CSPs) and the Tenant. This diagram shows two pathways for configuring GPU power profiles in data centers. The left path shows datacenter operators using out-of-band mechanisms to configure profiles across all nodes where a workload is running, for example to respond to a power demand response event, through NVIDIA Base Command Manager (BCM) Base View with Redfish APIs. The right path shows Tenants using workload scheduler plugins that communicate through in-band interfaces, such as, Nvidia-SMI or DCGM APIs. Both paths ultimately control the GPU driver to optimize power consumption based on workload requirements.

To facilitate these different access models, the infrastructure exposes a variety of APIs. For out-of-band management, an Administrator would utilize Redfish, an industry-standard API for hardware management, to apply configurations across the data center. This method is ideal for high-level, programmatic control. Conversely, a Tenant would use one of the several in-band APIs provided, such as NVSMI, NVML, DCGM, or BCM. These APIs offer a higher-level path through schedulers like SLURM and fine-grained control directly from the operating system, allowing a user to dynamically adjust settings based on the requirements of their specific workload, such as a particular LLM inference or training task.

Power profile selection integrates directly into standard job scheduler commands. An example of SLURM launch line is below for a MAX-Q-Training job:

sbatch --partition=gpu_partition --power-profile=MAX-Q-Training --nodes=4 --ntasks-per-node=8 training_job.slurm

Regardless of the API used, all configuration requests ultimately converge on the NVIDIA Kernel Mode Driver. This driver serves as the central control plane, where the core function of arbitration takes place. Arbitration logic is the "brain" of the infrastructure, responsible for resolving any potential conflicts or overlaps between different power profile configurations. By using predefined rules, priorities, and conflicting masks, the driver intelligently determines the final, singular configuration. This final, arbitrated configuration is then applied directly to the device (GPU), ensuring a stable and optimized state for the running workload

**Layer 4: Infrastructure Orchestration via NVIDIA Mission Control.** The final layer integrates this entire stack into a comprehensive infrastructure management solution: NVIDIA Mission Control. Mission Control is the first high level interface for user and administrator comprehensive configuration management and control enabling access to all layers of the stack from a single place. High level interfaces only available at layer 4 and lower-level interfaces, such as, the NVIDIA-SMI are exposed in Mission Control. Future plans include enabling other ways for users and administrators to access all layers and high-level interfaces from a single tool, with DCGM the next most likely path.

Mission Control serves as a higher-level orchestration platform providing a single comprehensive tool to access all the layers of the power profiles stack from a high-level interface. It also serves as a hub and coordination mechanism for other NVIDIA power tools enabling all reporting and configuration from the same place. The integration creates seamless workflows from job submission to performance monitoring while maintaining compatibility with existing datacenter management tools.

Mission Control simplifies the process of using profiles for administrators and end-users. It allows administrators to configure and deploy system-wide power profiles in coordination with other power control tools and monitoring capabilities. End-users, similarly, have their workflows simplified through a single point of entry. Mission Control enables site wide integration and provides real-time monitoring dashboards. It can be configured to automatically enforce site-wide power policies, including system wide power profiles, by integrating with facility-level building management systems (BMS) and hardware-level controllers.

The Mission Control integration spans all architectural layers:

**Layer 1 Integration:** Mission Control interfaces with foundational hardware and firmware controls through established APIs, providing administrators access to GPU and CPU clock controls, memory frequency settings, power limits, and NVLink power states without requiring deep understanding of complex hardware interdependencies.

**Layer 2 Integration:** The platform leverages the Profile Abstraction Layer's firmware tables and pre-configured profiles, enabling holistic power management rather than individual hardware control manipulation. This integration provides access to IP-level control algorithms and embedded optimization schemes that analyze workload characteristics automatically.

**Layer 3 Integration:** Mission Control extends Management and Monitoring APIs through higher-level orchestration capabilities while maintaining full compatibility with existing NVIDIA-SMI and DCGM interfaces, ensuring continued functionality of established datacenter tools and workflows.

**Layer 4 Implementation:** Mission Control provides the enterprise management layer with policy enforcement, facility integration, near real-time power consumption monitoring, and power profile selection based on workload characteristics and operational constraints.

## 3 Available Profiles and Advance Features

This section describes the power profiles that are currently available and those still in development. It also outlines the advanced support that mission control provides for job preparation and monitoring.

### 3.1 Available Power Profiles and Configuration

Workload Power Profiles currently provides four fully optimized, pre-configured, and validated profiles in its initial release with four additional profiles under development for scientific computing applications [10]. The four released profiles are:

- **Max-P Training**: Performance optimized for AI training workloads requiring maximum computational throughput.
- **Max-P Inference**: Performance optimized for AI inference workloads with latency-sensitive requirements.
- **Max-Q Training**: Energy efficiency optimized for AI training workloads. Enables best energy efficiency and maximizing throughput in power-constrained environments
- **Max-Q Inference**: Energy efficiency-optimized for AI inference workloads prioritizing efficiency and throughput in power constrained environment over peak performance.

The following profiles primarily for scientific computing applications are currently under development with initial results in the next section:

- **Max-P HPC-Compute**: Performance optimized for scientific computing workloads that heavily use FP32/FP64 compute.
- **Max-P HPC-Memory**: Performance optimized for memory bandwidth bound scientific computing workloads that use FP32/FP64 compute.
- **Max-Q HPC-Compute**: Energy efficiency optimized for scientific computing workloads that heavily use FP32/FP64 compute.
- **Max-Q HPC-Memory**: Energy efficiency optimized for memory bandwidth bound scientific computing workloads that use FP32/FP64 compute.

Each profile is generated by tuning the optimal recipe of following GPU configurations:

| Configuration | Impact |
| --- | --- |
| Total Graphic Power (TGP) | Dynamically adjusts GPU power limits based on the workload characteristics to optimize performance within power budgets. |
| EDP | Balances computational throughput with energy consumption to minimize total energy cost for completing workloads. |
| MCLK | Optimizes memory subsystem power consumption based on bandwidth requirements for specific applications. |
| XBAR:GPC | Adjusts interconnect power states to minimize communication overhead while maintaining computational throughput. |
| Fmax | Intelligently scale operating frequencies to match workload computational requirements without unnecessary power consumption. |
| Nvlink L1 | Optimizes high-speed GPU interconnect power states based on inter-GPU communication pattern. |
| RBM | Dynamically allocates GPU compute resources to maximize utilization efficiency while maintaining power consumption. |

These parameters work synergistically to achieve optimal performance-per-watt ratios. The TGP and Fmax settings provide coarse-grained power control, establishing fundamental power envelopes that define operational boundaries for different workload classes. Meanwhile, MCLK and NVLink optimizations address memory and communication bottlenecks that often limit real-world application performance, ensuring that power reductions don't create new performance constraints in critical data pathways.

The EDP optimization represents a sophisticated approach to energy management that considers both computational efficiency and temporal constraints. Rather than simply reducing power consumption, EDP ensures that power reductions translate to meaningful energy savings while maintaining acceptable job completion times. This prevents scenarios where reduced power leads to proportionally longer execution times, negating energy benefits.

## 3.2 Advanced Capabilities Exposed Through Mission Control

Mission Control provides advanced monitoring, analysis, system management, and workflow capabilities when using power profiles. Capabilities are exposed to both administrators and end users with plans in the future to expose these through other tools, such as DCGM.

Multi-level power monitoring and performance metrics are included. Monitoring tracks power and energy consumption from the individual GPU level through the node and rack level up to the whole facility providing granular visibility into energy usage patterns. The system as well as individual jobs are tracked, with metrics including computation throughput, energy efficiency and facility-level power utilization reported. Expected vs. actual power and energy savings are also reported. Meta-data, such as the profile enabled and application run when known are stored along with power and energy used. This enables historical analysis to aid future profile selection and optimize how the system is running. Power profiles will be enhanced in the future to use this data to help automate profile selection and provide suggestions to end users and administrators.

Administrator tools enable automated policy enforcement and modification. Through Mission Control datacenter administrators can set and modify power profile policies at the system of building level. This enables responding to external factors, such as, demand response events, power grid conditions, changing electrical costs, or cooling constraints. In addition, alerts can be configured to notify administrators when profile settings cause performance degradation to drop below a configured threshold.

User tools manage the job lifecycle from submission through performance analysis, ensuring optimal power efficiency throughout execution. Mission Control can provide power profile selection guidance based on workload characteristics, historical performance data, and current facility constraints. Upon job submission, it validates power profile compatibility with requested resources and available power budget. The system then tracks real-time performance metrics and performs post-execution analysis. This analysis quantifies performance impact, power savings, and throughput improvements and can provide recommendations for profile adjustments in future job submissions. Machine learning models are planned for future automated profile selection capabilities.

## 4 Performance Benefit Evaluation

The experiments shown in the paper data were collected on two types of systems. Both systems are NVIDIA HGX based systems each with eight GPUs connected together by NVLINK per node and two CPUs connected to the GPUs via PCIe. Most of the results shown were collected on the Blackwell B200 system, which has 1000W NVIDIA Blackwell GPUs and 56 core Intel Emeral Rapids 8570 Xeon CPUs. The other system used contains 700W NVIDIA H100 GPUs and 56 core Intel Saphire Rapids 8480C CPUs. In both machines the nodes are connected with a NDR 400 Gb/s network with one network interface connection per GPU. Unless noted tests were run on 8 to 12 nodes.

All baseline data shown was collected with default settings without enabling workload power profiles. For all runs application performance and power of all components in the system were collected. Most power profile runs used the Max-Q profiles to see the performance and power impact of using workload power profiles. We include some Max-P data for completeness. The primary initial use case is energy efficiency and the Max-Q mode today. However, over time we expect as systems become more power constrained, even at TDP, that Max-P will be used more heavily. For example, NVIDIA MLPerf solutions already are see performance improvements from power settings when running at TDP [5].

Profiles are application class specific because they use the hardware differently. For example, AI heavily uses tensor cores and 16 bit or lower precisions, while HPC rarely uses tensor cores and is mostly FP32 or FP64. This leads to applications using hardware differently. For example, these two applications are often running at different points of the VF curve. This is why the tool performs best when given both application type and hints about the code e.g. memory vs compute bound. In future these will be self-detected based on hardware telemetry that expose characteristics of applications.

The following set of data across B200 and H100 shows the effectiveness of Max-Q and Max-P profiles across AI and HPC applications. As AI datacenters are becoming power constrained, our key focus has been on improving Datacenter throughout by improving System Energy efficiency and hence datasets show the benefits of deploying Max-Q profiles across applications.

| Application | Performance Loss | Data Center Power Saving | Data Center Throughput Increase |
|---|---|---|---|
| DeepSeek R1 | 3% | 12% | 8% |
| Llama 3.1 8B | 2% | 11% | 7% |
| Llama 3.1 70B | 2% | 9% | 6% |
| Mistral 7B | 2% | 9% | 6% |
| HPL | 1% | 13% | 12% |
| GROMACS | 1% | 15% | 13% |
| LAMMPS | 2% | 14% | 13% |
| RTM | 2% | 13% | 12% |

Table I: Effects of Applying Max-Q Profiles to Various AI and HPC Applications.

TABLE I. shows the impact of running Max-Q power profiles on a mix of AI and HPC applications with a performance loss threshold of 3%. The first column lists the application, the second shows performance loss, the third indicates power saved, and the fourth shows the increase in power-constrained data center throughput using Max-Q mode as power profiles enable to fit more GPUs into a power constrained Datacenter.

The data in TABLE I. has two clusters for the data center throughput increase metric. The first four applications are AI and have a data center throughput of 6-8% while the last four applications, which are HPC, have a throughput increase of 12-13%. The difference in throughput increase is due to a few factors. The HPC applications, even though three of the four are compute limited, are running at a higher clock frequency so scaling down clock frequency to save power leads to larger power savings. Most of the HPC apps don't consume full TGP and hence percentage of saved power to total is higher compared to AI apps. These applications do not use high power interconnects such as NVLINK unlike the AI applications thus enabling larger savings.

| Training Applications | GPU Power Savings | System Power Saving | Job Energy Savings |
|---|---|---|---|
| NeMo_gpt3_5b | 4% | 8% | 7% |
| NeMo_llama3_8b | 5% | 8% | 6% |
| NeMo_nemotron_22b | 18% | 12% | 10% |
| PyTorch_bert_large | 16% | 10% | 8% |

Table II: Max-Q GPU and system power savings and job energy savings for training applications on B200

TABLE II. shows the energy efficiency savings of the Max-Q profiles across additional training jobs. In this example we break out the power savings from the GPU and at the system level separately. GPU savings come from lower clock frequencies and by reducing power to underutilized structures within the GPU, such as, the crossbar. While GPU power savings are highly varied among these models with a wide range from 4-18% job power savings are similar as the applications with `larger power savings slowed down more than the ones with smaller power savings.

**Table III: Max-Q System perf/watt and Energy efficiency Training across key applications on B200**

| Metrics | AI Apps | HPC Apps |
|---|---|---|
| Performance | -2% | -1% |
| GPU power savings | 11% | 13% |
| System power savings | 9.5% | 11% |

Table III breaks out detailed properties as an average for the AI and HPC applications in Table I. It shows the average statistics from the previous table for perf on average, across training workloads, Max-Q profiles on average achieve about a 10% system power savings. These GPU power savings translate to the system since other components outside the GPU also scale with these settings hence saving system power and perf/watt almost in line with GPU savings.

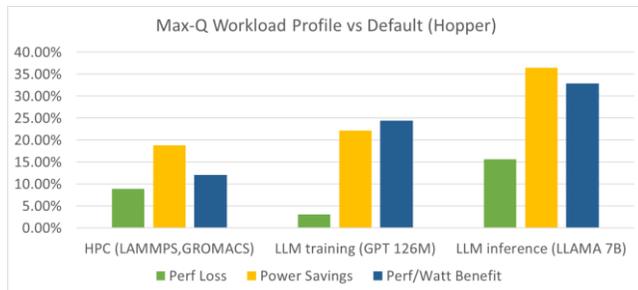

**Figure 3: Max-Q perf, power and perf/watt comparison across AI and HPC workloads on Hopper**

Fig 3 shows the Max-Q profile benefits for Energy efficiency across HPC and AI workloads on Hopper when performance loss is uncapped. Power savings are range from 18-36%, but come at the expense of a 3-16% performance decreases. Overall, energy savings (perf/W) improves 12-32%. With uncapped performance losses power and energy savings can be significantly increased verse our previous experiments.

In contrast to Blackwell AI applications have larger power savings and similar performance losses as HPC applications. This switch is because HPC applications scale with SMs. Hopper has 13% fewer SMs vs. Blackwell while using 30% less power indicating that there is less inefficiently used power for HPC applications on Hopper as power per SM is lower. For AI by contrast Hopper has 60% less tensor core compute so on Hopper AI applications are running at a less efficient point of the voltage frequency curve enabling more power savings opportunities.

**Table IV: Comparison of Performance Drops from Frequency Scaling to Save Power Vs. Using Power Profiles**

Table IV compares scaling GPU clock frequency only to save power verse using power profiles and the resulting performance impact. We average over the inference and training results presented previously in this section. Since power profiles result in

| Blackwell B200 | Performance Decrease | Data Center Power Savings |
|---|---|---|
| Frequency Scaling | 10% | 5% |
| Training Profiles | 1% | 5% |
| Inference Profiles | 3% | 8% |

a minimum 5% power savings we set this as the average goal for frequency scaling. Our results show that power profiles see a 7-9% smaller performance decrease and the same or larger data center power savings as frequency scaling. Frequency scaling only has a large impact on performance because the 1000 Watt B200 part is operating at an efficient point on the voltage frequency curve and the power of the rest of the system is unchanged. Power profiles by contrast saves power in other parts of the system that impact performance significantly less for workloads that are compute sensitive.

The results in this section across H100 and B200 show that across key AI training, inference and HPC workloads, the Max-Q profiles achieve significant energy savings over default settings. Historically, this has taken many engineering months for users to tune individual workloads with usage mostly limited to experts in the research community. With power profiles any user can get out of the box an energy efficient profile for their workloads without spending any effort on tuning for their workloads.

**Max-P Profiles:** Like Max-Q profile, Nvidia also provides out of the box Max-P profiles. Max-P profiles aim to improve performance by lowering the performance of hardware components that do not limit performance and moving it to others where it can be used to boost application peak performance. The performance boost for compute bound workloads is typically achieved by saving power on Nvlink, PCIe and Memory for workloads that don't utilize Memory/IOs and diverting it towards the GPCs. Memory-bound workloads typically do not benefit from Max-P mode as they usually, but not always, run below TDP on Hopper and Blackwell architectures.

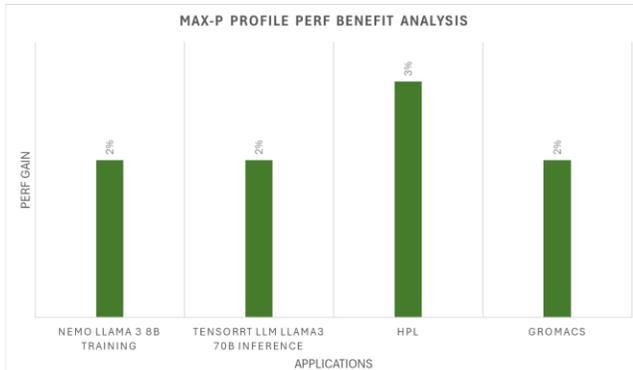

Figure 4: Max-P Performance improvement from power profiles for a mix of HPC and AI workloads

Fig 4 shows Max-P profile benefit across a mix of HPC and AI applications on B200. Overall, we can observe 2-3% performance gain in Max-P. Gains are smaller for Max-P because the GPU is typically running at a less efficient point in the voltage frequency curve already, so significant power is needed to boost performance. In addition, while the throttled structures use significant power they cannot be fully turned off and they use less than the compute units and caches.

## 5 Related Work

Power and energy optimization in running systems is becoming increasingly common as power, both usage and provisioning is a large portion of the total cost of ownership of a supercomputer. A breadth of techniques is being applied throughout German supercomputing sites, including, warm water cooling, waste heat recapture, and using job specific insights [17]. HPE and HLRS collaborated on reducing the provisioning costs of power through intelligent power scheduling [15]. Both approaches are complementary to our work on running jobs more efficiently.

Initial work on power tuning focused on frequency settings. Lowering CPU frequencies for memory bound applications can save significant power with little performance loss c. Loadleveler predicted power frequency setting and showed sometimes increasing frequency to lower time to solution optimized energy usage [1]. These insights are included in our work, which also touches many other settings.

Various tools enable runtime setting of power including EAR [3] and GeoPM [5]. EAR is used in production and runs outside the job enabling an increase in usability and influences clock frequencies. Similar to EAR we reduce the burden on the use but leverage a small amount of additional user provided data to make more informed decisions. GeoPM enables significant control but requires more detailed user analysis to use than our tool. We target static settings with our approach that enables modification of settings, such as HBM memory frequency that cannot be set at runtime, and more settings than most users will adjust on their own. In addition, tools like GeoPM can interact with jobs using our settings to adjust some settings at runtime to further improve energy efficiency.

Our approach is complimentary to work on scheduling overprovisioned systems [13][15]. By enabling applications to run as efficiently as possible at different user or administrator defined power levels with minimal effort we make the practical challenges of using overprovisioned systems easier. When power is available, users can take advantage of it and use Max-P mode, and when its not available they can use Max-Q to fit into their individual power budget. Reducing idle power usage is another way to save energy [9] that is similar to our approach. Most of our power savings come from reducing power to idle or underused structures at runtime.

Most similar to our work of using high level input to set fine-grain power control is [16] which also allows high level selection of policies to optimize energy to solution on the CPU Fugaku supercomputer. Both approaches set static settings at job launch. However, the Fugaku approach has users selecting their own power setting while instead we enable users to describe their applications and goals and then suggest the best policy for their job. This reduces the amount of expertise a user needs to effectively use our tool.

In addition, to power tools recent work has identified significant underutilized power in data centers. Many applications, especially memory bound ones run significantly below TDP[2]. To use this power overprovisioning has been proposed and analyzed [14]. Worries over power grids hitting their limits are leading to demand response agreements to access grid power [15]. Our work is complimentary with these approaches and observations as it enables more efficient use of power, which maximizes output from this limited resource.

## 6 Future Roadmap

Figure 5 shows our roadmap for the power profile tool. This paper presents production results for gen-1 of this feature and initial results for gen-2, which is efficient GPU power profiles for AI and HPC applications on Blackwell product lines. The out of box solution delivers fully optimized recipes in form of power profiles encapsulating GPU DVFS, performance and power tunable knobs, considering GPU design architecture and constraints, software / firmware control algorithms, inter-dependencies of configuration and workload awareness from our chip post-silicon validation process.

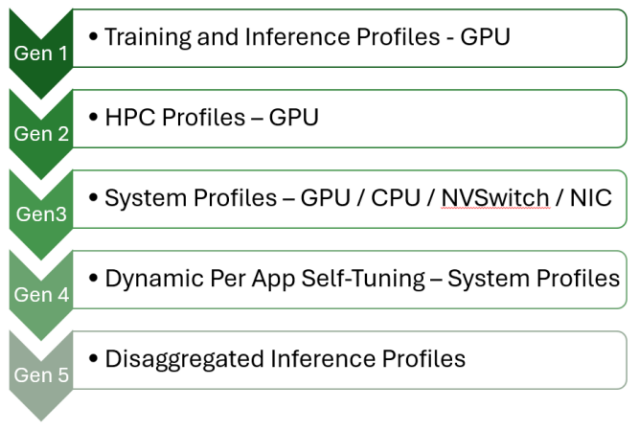

**Figure 5: Power Profiles Roadmap**

Our vision is to bring energy efficiency across the Datacenter stack by delivering fully optimized recipes in the form of power profiles, encapsulating optimization at all the layers starting from ground level hardware IPs (CPU, GPU, IO/Networking) with their software / firmware control systems, Datacenter form factors compute trays, servers, rack, Datacenter operating conditions Power and Thermal constrains, and Datacenter workload / applications.

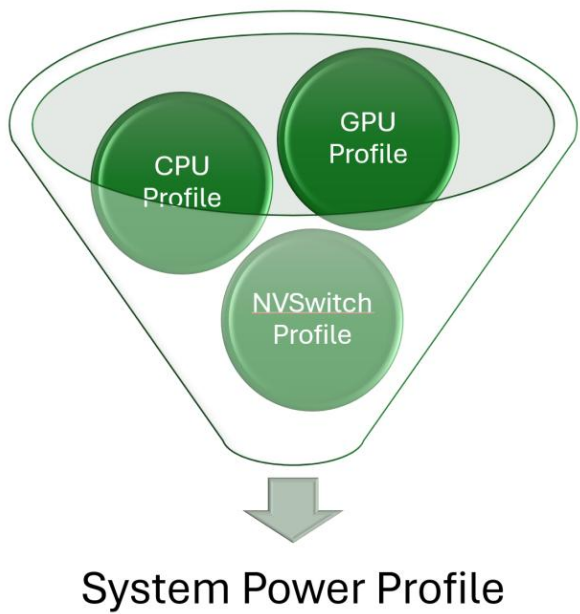

**Figure 6: System Power Profiles Composition**

Once our second-generation product incorporating HPC application power profiles for GPUs is done we will begin work on the third generation product expanding power profiles to incorporate tunable configurations from CPU, networking, NVLINK, InfiniBand, and Ethernet, in addition to other system components as appropriate to deliver system power profiles Fig 6. These profiles will leverage our knowledge of our hardware architecture design, core firmware algorithms such as DVFS, and performance and power management. We are also looking into reducing stranded power in the system through system profiles enforcing power limits on devices like GPU, CPU and NVSwitch leveraging our data analysis of AI and HPC workloads, optimizing performance at watt. Additionally, work is ongoing for improving the efficacy of current profiles via additional optimization hints from users around properties of a workload, such as compute-bound, memory-bound or I/O limited. Nvidia Mission Control integration will enable Datacenter administrators to select appropriate system profile taking into account datacenter operating conditions and job characteristics such as complexity, priority, etc. The selected system profile will enforce optimal device specific profiles in addition to system level tuning for desired behavior.

Our fourth-generation product plans to enable further workload specialization beyond user hints. Datacenter use cases are rapidly evolving and while static configuration based on offline data analysis provides significant improvements from the baseline, the optimal energy efficiency can only be reached through dynamic per application tuning. We plan to accomplish this by leveraging live telemetry from devices while an application runs. Then the device configuration will adapt to the active workload. The team is performing multiple paths to enable this, including investment into deep learning where neural networks are trained offline with AI and HPC use cases to provide hints to DL control algorithms for efficiently navigating various tunable knobs and/or system profiles, delivering desired peak performance/energy efficiency goals of application users and datacenter administrators. Finally, in the last planned phase of our roadmap we will incorporate support for disaggregated inference.

Throughout this process, we will continue to add additional HW and SW settings that our software can modify as future GPUs/CPUs/Networking chips provide more advanced options for power control. We anticipate the number of control knobs will grow exponentially from a handful (<10 in Blackwell) as systems become more power constraint and the ability to optimize power and energy becomes more important. The growing importance of tuning the power of other system components will only exacerbate this trend. We believe only a profile infrastructure can abstract the growing number of settings and enable tuning for specific applications without heroic user effort.

## 6 Conclusion

In this paper, we have introduced the power profiles feature released with NVIDIA Blackwell GPUs. Power profiles minimize programmer effort to optimize multiple settings concurrently to increase energy efficiency by providing a high level easy to understand interface. This eliminates the need for users to adjust these settings individually significantly reducing the effort to optimize power and energy. On Blackwell our implementation achieves up to 15% energy savings with at most a 3% performance loss for various HPC and AI applications resulting in an overall throughput increase of up to 13% in a power-constrained facility. On Hopper with no maximum performance loss energy efficiency improves up to 32%. In addition, the maximum performance mode improves application performance by up to 10%.

We also present our roadmap for further improving the capabilities of power profiles. Future features include extending the capabilities to the whole system (NICs, switches, etc.) and handling disaggregated inference applications. In addition, we plan to add the ability for users to describe their application bottlenecks and profiling and AI tools to help automate profile selection and further ease both use and effectiveness of this feature for non-expert users.